\documentclass[a4paper,11pt]{article}
\pdfoutput=1 % if your are submitting a pdflatex (i.e. if you have
\usepackage{jcappub}

\title{Inflationary cosmology with Chaplygin gas in Palatini formalism}

\author[a,d]{Andrzej Borowiec}
\author[b]{Aleksander Stachowski}
\author[b,c]{Marek Szyd{\l}owski}
\author[a,e]{Aneta Wojnar}

\affiliation[a]{Institute for Theoretical Physics, University of Wroc{\l}aw, pl. M. Borna 9, 50-204, Wroc{\l}aw, Poland}
\affiliation[b]{Astronomical Observatory, Jagiellonian University, Orla 171, 30-244 Krak{\'o}w, Poland}
\affiliation[c]{Mark Kac Complex Systems Research Centre, Jagiellonian University, ul. {\L}ojasiewicza 11, 30-348 Krak{\'o}w, Poland}
\affiliation[d]{Dipartimento di Scienze e Innovazione Tecnologica
Universit`a del Piemonte Orientale
Viale T. Michel 11, 15121 Alessandria, Italy}
\affiliation[e]{Istituto Nazionale di Fisica Nucleare (INFN) Sez. di Napoli, Compl. Univ. di Monte S. Angelo,
Edificio G, Via Cinthia, I-⁠80126 Napoli, Italy}

\emailAdd{andrzej.borowiec@ift.uni.wroc.pl}
\emailAdd{aleksander.stachowski@uj.edu.pl}
\emailAdd{marek.szydlowski@uj.edu.pl}
\emailAdd{aneta.wojnar@ift.uni.wroc.pl}

 \abstract{
We present a simple generalisation  of the $\Lambda$CDM model which on the one hand reaches very good agreement with the present day experimental data and provides an internal inflationary mechanism on the other hand. It is based on  Palatini modified gravity with quadratic Starobinsky term and generalized Chaplygin gas as a matter source providing,  besides a current accelerated expansion, the epoch of endogenous inflation driven by type III freeze singularity. It follows from our statistical analysis that astronomical data favors negative value of the parameter coupling quadratic term into Einstein-Hilbert Lagrangian and as a consequence the bounce instead of initial Big-Bang singularity is preferred.}

\begin{document}
\maketitle
\flushbottom

\section{Introduction}
Extended Theories of Gravity (ETG's) have been gained a lot of interest as some of them could help us to understand shortcomings in theoretical
physics which General Relativity (GR) \cite{Einstein:1915ca,Einstein:1916vd} is
not able to explain (for reviews, see e.g. \cite{Copeland:2006wr,Nojiri:2006ri,Capozziello:2007ec,Carroll:2004de,Sotiriou:2008ve}). One tries to modify the geometric part of field equations (it means the gravity action) as well as the matter part in order to clarify problems like inflation \cite{Starobinsky:1980te,Guth:1980zm}, dark energy \cite{Copeland:2006wr,Nojiri:2006ri,Capozziello:2007ec,Huterer:1998qv}, dark matter, large scale structure. Moreover, one would like to reveal a way to formulate Quantum Gravity. The first modification of Einstein's field equations was done by Einstein himself %\textcolor{red}{cytacja?}
by introducing a cosmological constant to the gravitational action in order to make the
cosmological solutions static. Adding to the action scalar fields as an additional dynamical variable, considering exotic fluids instead of standard baryonic matter, generalizing the action functional, varying physical constants (velocity of light, gravitational constant) and many others
\cite{Copeland:2006wr,Nojiri:2006ri,Capozziello:2007ec,Capozziello:2010be} %\textcolor{red}{cytacje, chronologia}
have been considering theoretically and experimentally by many authors, although no theory has been set as appropriate so far. There is still room for new investigations of the existing theories as many of them pass the solar system tests \cite{DeFelice:2010aj,Will:1993te}. Each theory must be confronted with the newest observational data \cite{Suzuki:2011hu,Outram:2003ew,Paris:2012iw,Schneider:2010hm} and should be compared with the standard cosmological model, so-called $\Lambda$CDM (see e.g. \cite{Borowiec:2006hk,Borowiec:2006qr,Borowiec:2011wd,Borowiec:2013ima}).

In this paper we have proposed two modifications of the standard theory of gravity. The first correction concerns the geometric part of the field equations; we have studied $f(\hat{R})$-gravity in Palatini formalism \cite{Palatini:1919di,Sotiriou:2008rp,Capozziello:2011et,Ferraris:1992dx}. In that approach the torsionless connection $\hat{\Gamma}$ is treated as a variable independent of the spacetime metric $g_{\mu\nu}$ and it is used to construct Riemann and Ricci tensor. The approach possesses a huge advantage in comparison to $f(R)$-gravity considered in the metric formalism \cite{Sotiriou:2008rp,Carroll:2004de,Sotiriou:2008ve,DeFelice:2010aj,Will:1993te} since the equations of motion are second order ones (in $f(R)$-gravity in the metric formalism one deals with fourth-order differential equations). In the case of GR, it means when one considers the Einstein - Hilbert Lagrangian $f(\hat{R})=\hat{R}$, the variation with respect to the connection arises to a conclusion that the connection is the Levi-Civita connection of the metric $g_{\mu\nu}$. But the situation is different for ETG's, that is, for more complicated functions $f(\hat{R})$.

Although the $f(\hat{R})$-gravity in the Palatini approach suffers problematic issues like being in conflict with the Standard Model of particle physics \cite{Sotiriou:2008rp,Flanagan:2003rb,Iglesias:2007nv,Olmo:2008ye}, surface singularities of static spherically symmetric objects in the case of polytropic equation of state \cite{Barausse:2007ys}, the algebraic dependence of the post-Newtonian metric on the density \cite{Olmo:2005zr,Sotiriou:2005xe}, and the complications with the initial values problem in the presence of matter \cite{Sotiriou:2005xe,Sotiriou:2006hs}, it is still a subject of many investigations in cosmological
applications \cite{Fan:2015lta,Capozziello:2015wsa} %\textcolor{red}{cytacji wiecej}
astrophysics \cite{Olmo:2015xwa,Olmo:2015bya,Bambi:2015zch} and others \cite{Briscese:2015urz,Bambi:2015sla,Guendelman:2012ve}. There are also works on hybrid metric-Palatini gravity that combines two formalisms: metric and Palatini \cite{Harko:2011nh,Capozziello:2012ny,Borowiec:2014wva}. It is also worth to mention that Palatini formalism is widely used in such theories like e.g. $f(\hat{R}_{\mu\nu}\hat{R}^{\mu\nu})$-gravity (see e.g. \cite{Carroll:2004de,Sotiriou:2008ve} for review) because it cures the complication with the connection treated as an auxiliary field promoting it to a truly dynamical field \cite{Barausse:2007ys,Li:2007xw} without making the equations of motion second order in the fields. In the metric approach one deals with a problem called the Ostrogradski instability \cite{Woodard:2007ad}.

The second modification is applied to the equation of state (EoS) $p=p(\rho)$. In cosmology, in the standard approach one uses the barotropic EoS $p=\omega\rho$, where $\omega$ is a dimensionless number indicating a kind of matter. In our work we would like to investigate exotic fluid so-called Generalized Chaplygin Gas (GCG) \cite{Bento:2002ps,Lima:2006qe} which recently has gained a lot of attention in the literature \cite{Bazeia:1998tf,Jackiw:1999wy,Ogawa:2000gj}. The interesting feature of Chaplygin gas is that it is the only fluid known up to now which has a supersymmetric generalization \cite{Hoppe:1993gz,Jackiw:2000cc}. Moreover, it has also a representation as tachyon field \cite{Benaoum:2002zs,Chimento:2003ta}. It was introduced in order to compute the lifting force on a wing of an airplane in aerodynamics \cite{Chaplygin:1904cg} but it has also been used in cosmology \cite{Bento:2002ps,Kamenshchik:2001cp,Lu:2009zzm,Bilic:2001cg,Popov:2009bt,Naji:2014kaa,Kremer:2004bf,Gorini:2002kf,Avelino:2014nva,Kahya:2015dpa,Fabris:2010vd} since it can be seen as realization the idea of quintessence for unification both dark mater and dark energy.

Concluding, we shall consider the homogeneous and isotropic universe filled by the GCG and baryonic matter. We will investigate cosmological models of extended cosmology basing on generalization of GR to $f(R)= R+\gamma R^2$ theory of gravity \cite{Starobinsky:1980te}. Our choice of the Lagrangian is motivated by a possibility of an explanation different problems in modern cosmology like for example the problem of
sterile neutrinos, the problem of acceleration of the current Universe and others already mentioned. These $f(R)$ models offer the explanation of present dark
energy \cite{Starobinsky:2007hu} and satisfy all existing observational constraints. They are interesting in the context of study of the deviation of
the background evolution from the standard $\Lambda$CDM model which is very small, less than a few percent (see \cite{Motohashi:2010tb,Motohashi:2012wc} and see for recent results \cite{Ade:2013zuv}).

For $f(R)$ cosmologies there is a very interesting approach which allows to explain the inflation as an endogenous effect. One considers such ideas like intermediate inflation (for discussion of intermediate inflation see also \cite{Herrera:2013ust,Herrera:2013rca}) or inflation driven by type IV singularities \cite{Nojiri:2015wsa,Nojiri:2005sx}.

In our work, we shall study observational constraints on the cosmological model $f(\hat{R})=\hat{R}+\gamma\hat{R}^2$ in the Palatini formalism. We will estimate, using the
recent astronomical data, model parameters and compare the model with the standard cosmological one $\Lambda$CDM. We will incorporate the idea of quintessence by considering the GCG as a source of gravity. Additionally, the visible baryonic matter will be added to the matter part.

From our statistical analysis it will be shown  that the model under consideration arises to examine the effect which might be
treated as a small deviation from the $\Lambda$CDM model (represented later by the density parameter $\Omega_\gamma$). It introduces a freeze type singularity
in the early evolution of the Universe. The correspondence with the standard cosmological model is achieved
if the deviation term goes to zero. We calculate the likelihood interval for value of the deviation parameter $\Omega_{\gamma}$ which
belongs to the interval [0, $10^{-9}$].

In the estimation of the model parameters we have obtained an exact formula for the Friedmann equation $H^2$ which is written as a function of redshift. Moreover, for the special case of the pure
Chaplygin Gas we have also received an exact formula for a value of redshift corresponding to a freeze singularity as a function of model parameters.

The paper is written as follows. In the section \ref{prelim} we will give some basis of the Palatini cosmology and GCG. In the section \ref{par} we will
introduce a suitable parametrization of the model and perform analytic examination. The section \ref{par} will treat about the statistical analysis
and the results will be presented. We will shortly conclude our investigation in the section \ref{conclus}.

\section{Preliminaries}\label{prelim}
Let us shortly discuss the formalism that we are going to consider in the paper. The action of the Palatini $f(R)$ theories
is of the form:
\begin{equation}\label{action}
 S=\frac{1}{2\kappa}\int\mathrm{d}^4x\sqrt{-g}f(\hat{R})+ S_m,
\end{equation}
where $f(\hat{R})$ is a function of a Ricci scalar $\hat{R}=g^{\mu\nu}\hat{R}_{\mu\nu}(\hat{\Gamma})$ constructed by
the metric-independent torsion-free connection $\hat{\Gamma}$ and $S_m$ is a matter action
independent of the connection. One varies the action with respect to the metric and obtains
\begin{equation}\label{var_met}
 f'(\hat{R})\hat{R}_{(\mu\nu)}-\frac{1}{2}f(\hat{R})g_{\mu\nu}=\kappa T_{\mu\nu}.
\end{equation}
The prime denotes differentiation with respect to $\hat{R}$ and $T_{\mu\nu}$ is the standard (symmetric) energy-momentum
tensor given by the variation of the matter action with respect to $g_{\mu\nu}$. The $g$-trace
of (\ref{var_met}) gives us the structural equation of the spacetime controlling (\ref{var_met})
\begin{equation}\label{struc}
  f'(\hat{R})\hat{R}-2f(\hat{R})=\kappa T.
\end{equation}
Assuming that we can solve (\ref{struc}) as $\hat{R}(T)$ we get that $f(\hat{R})$ is a function
of $T$, where $T=g^{\mu\nu}T_{\mu\nu}$ is a trace of the energy-momentum tensor.
The variation with respect to the connection can be written as
\begin{equation}
 \hat{\nabla}_\lambda(\sqrt{-g}f'(\hat{R})g^{\mu\nu})=0
\end{equation}
from which it follows that the connection is the Levi-Civita connection for the conformally related
metric $f'(\hat{R})g_{\mu\nu}.$ For more detailed discussion see for
example \cite{Allemandi:2005qs,Allemandi:2004ca,Allemandi:2004wn}.

In the following paper we will consider perfect fluid energy-momentum tensor
\begin{equation}\label{perfect}
 T_{\mu\nu}=\rho u_\mu u_\nu + p h_{\mu\nu},
\end{equation}
where $\rho$ and $p=p(\rho)$ are energy density and pressure of the fluid, respectively. The vector $u^\mu$ is an observer
co-moving with the fluid satisfying $g_{\mu\nu}u^\mu u^\nu=-1$ and $h^\mu_\nu=\delta^\mu_\nu+u^\mu u_\nu$ is
a 3-projector tensor projecting 4-dimensional object on 3-dimensional hypersurface in the case when the observer $u$ is rotation-free.
Hence, the trace of (\ref{perfect}) is
\begin{equation}\label{trace}
 T=3p-\rho.
\end{equation}
Following the approach of \cite{Allemandi:2005qs,Allemandi:2004ca,Allemandi:2004wn}, the generalized Einstein's equations (\ref{var_met}) can be written as
\begin{equation}\label{EinR}
 \hat{R}_{\mu\nu}(\Gamma)=g_{\mu\alpha}P^\alpha_\nu,
\end{equation}
where the operator $P^\alpha_\nu$ consists of two scalars $b=f^\prime (\hat{R})$ and $c=\frac{1}{2} f(\hat{R})$ depending on $\hat{R}$
\begin{equation}\label{operator}
 P^\alpha_\nu=\frac{c}{b}\delta^\mu_\nu+\frac{1}{b}T^\mu_\nu,
\end{equation}

The background metric under our consideration is the Friedmann-Robertson-Lemaitre-Walker (FRLW) metric
\begin{equation}\label{frlw}
 ds^2=-dt^2+a^2(t)\left[ \frac{1}{1-kr^2}dr^2+r^2(d\theta^2+\sin^2{\theta}d\phi^2) \right],
\end{equation}
where $k=0,1,-1$ is the space curvature and $a(t)$ is a scale factor depending on cosmological time $t$. The energy-momentum
tensor (\ref{perfect}) satisfies the metric covariant conservation law $\nabla^\mu T_{\mu\nu}=0$ as we are considering the
Palatini $f(\hat{R})$ gravity as a metric theory \cite{Sotiriou:2008rp}. Due to that fact, one deals with the continuity
equation
\begin{equation}\label{continuity}
 \dot{\rho}+3H(p+\rho)=0,
\end{equation}
where $H=\frac{\dot{a}}{a}$ is the Hubble constant. If one know the relation between the pressure $p$ and energy density $\rho$, i.e.
equation of state $p=p(\rho)$,  then
using (\ref{frlw}) and (\ref{continuity}), one may obtain a dependence $\rho$ on the scale factor $a(t)$.
The generalised Einstein's field equations (\ref{EinR}) for the FRLW metric is the generalised Friedmann
equation \cite{Allemandi:2005qs,Allemandi:2004ca,Allemandi:2004wn}
\begin{equation}
 \left( \frac{\dot{a}}{a}+\frac{\dot{b}}{2b} \right)^2 + \frac{k}{a^2}=\frac{1}{2}P^1_1-\frac{1}{6}P^0_0,
\end{equation}
where $P^1_1$ and $P^0_0$ are the components of (\ref{operator}). Therefore, the modified Friedmann equation
has the form \cite{Allemandi:2005qs,Allemandi:2004ca,Allemandi:2004wn,Borowiec:2011wd}:
\begin{equation}\label{fried}
 \left( \frac{\dot{a}}{a}+\frac{\dot{b}}{2b} \right)^2+\frac{k}{a^2}=\frac{1}{2}\left( \frac{c}{b}+\frac{p}{b} \right) -
 \frac{1}{6}\left( \frac{c}{b}-\frac{\rho}{b} \right).
\end{equation}

Let us draw our attention on the matter part of the modified Einstein's field equations. The energy-momentum tensor has a perfect
fluid form with the equation of state $p=p(\rho)$. In our work we will consider an exotic perfect fluid, the so-called
Generalized Chaplygin Gas (GCG) \cite{Bento:2002ps,Bertolami:2006zg}, that is, the equation of state is
\begin{equation}
 p=-\frac{A}{\rho^\alpha},
\end{equation}
where constants $A$ and $\alpha$ satisfy $A>0$ and $0<\alpha\leq1$. For $\alpha=1$ one deals with the original Chaplygin gas
\cite{Chaplygin:1904cg,Kamenshchik:2001cp}. Substituting GCG into the conservation law (\ref{continuity}) in FRLW spacetime we get
\begin{equation}
 \rho=\left(A+\frac{B}{a^{3(1+\alpha)}}\right)^{\frac{1}{1+\alpha}},
\end{equation}
where $B>0$ is an integration constant. One notices that in the GCG model the early stage of the Universe is dominated by dust
($\rho\propto a^{-3}$) while at late times by cosmological constant (vacuum energy, $\rho\simeq\text{const}$).

\section{Physical parameterization of Generalized Chaplygin Gas}\label{par}

Let us consider the ansatz $f(\hat{R})=\hat{R}+\gamma\hat{R}^2$ and apply it to the structural equation of the theory (\ref{struc}). With the help of the
equation (\ref{trace}) one gets the Palatini curvature scalar as a function of the
scale factor $a(t)$ for GCG
\begin{equation}
 \hat{R}=\left( A+Ba^{-3(1+\alpha)} \right)^{\frac{-\alpha}{1+\alpha}}\left(4A+Ba^{-3(1+\alpha)}   \right). \label{palatinir}
\end{equation}

For the further analysis, instead of the original $A$, $B$ parameters, it would be convenient to use
observational constraints on the parameters of GCG model which are strictly related with
physics \cite{Bento:2002ps}. For this aim let us introduce $A_{\text{s}}$ and $\rho_{\text{ch},0}$ defined as follows
\begin{equation}
	\rho_{\text{ch}}=\rho_{\text{ch},0}\left(A_{\text{s}}+
	\frac{1-A_{\text{s}}}{a^{3(1+\alpha)}}\right)^{\frac{1}{1+\alpha}}=
	3H_0^2\Omega_{\text{ch},0}\left(A_{\text{s}}+\frac{1-A_{\text{s}}}{a^{3(1+\alpha)}}\right)^{\frac{1}{1+\alpha}},
\end{equation}
where $A_{\text{s}} \rho_{\text{ch},0}^{1+\alpha}=A$, $\rho_{\text{ch},0}^{1+\alpha}(1-A_{\text{s}})=B$.
The quantities labeled by index `0' correspond to the present epoch.
We also introduce the quantity $K$, which will be useful later, as follows
\begin{equation}
	K=\frac{3A_{\text{s}}}{A_{\text{s}}+(1-A_{\text{s}})a^{-3(1+\alpha)}}.
\end{equation}
Since $a\in [0,\text{ }+\infty)$, $K(a)$ function is bounded $K\in[0,\text{ }3)$. It is related
to the squared velocity of sound
$c_s^2=\frac{\partial p}{\partial \rho}=\frac{\alpha A_{\text{s}}}{A_{\text{s}}+(1-A_{\text{s}})a^{-3(1+\alpha)}}=\frac{1}{3}\alpha K$.

Let us define a new dimensionless parameter related with the $\gamma$ parameter
\begin{equation}
\Omega_{\gamma}=3\gamma H_0^2,
\end{equation}
and new dimensionless functions related to the functions $\hat {R}$, $b$, $c$, $\rho_{\text{ch}}$ in the following way:
\begin{align}
 \Omega_R=&\frac{\hat{R}}{3 H_0^2}=\Omega_{\text{ch},0}\left(A_{\text{s}}+
 \frac{1-A_{\text{s}}}{a^{3(1+\alpha)}}\right)^{\frac{1}{1+\alpha}}\frac{4+ \frac{(1-A_{\text{s}})}{A_{\text{s}}}a^{-3(1+\alpha)}}{1+\frac{(1-A_{\text{s}})}{A_{\text{s}}}a^{-3(1+\alpha)}}=
\Omega_{\text{ch}}(K+1),\\
b=&f'(\hat{R})=1+2\Omega_{\gamma} \Omega_R=1+2\Omega_{\gamma}\Omega_{\text{ch}}(K+1),\\
\Omega_{\text{ch}}=&\Omega_{\text{ch},0}\left(A_{\text{s}}+\frac{1-A_{\text{s}}}{a^{3(1+\alpha)}}\right)^{\frac{1}{1+\alpha}}=
\Omega_{\text{ch},0}\left(\frac{3A_\text{s}}{K}\right)^{\frac{1}{1+\alpha}},\\
\Omega_c=&\frac{c}{3H_0^2}=\frac{1}{6H_0^2}f(\hat{R})=\frac{\hat{R}}{6H_0^2}(1+\gamma\hat{R})=\frac{\Omega_{R}}{2}(1+\Omega_{\gamma}\Omega_{R})\nonumber\\
=&\frac{\Omega_{\text{ch}}(K+1)}{2}(1+\Omega_{\gamma}\Omega_{\text{ch}} (K+1)).
\end{align}
Note that $d(t)\equiv\frac{1}{H}\dot{b},\;\dot{b}=\frac{db}{dt}$ function can be rewritten in the terms of $K$
\begin{equation}
d=2\Omega_{\gamma}\Omega_{\text{ch}}(3-K)[\alpha(1-K)-1].
\end{equation}

Including to the model spatial curvature effects, the normalized Friedmann equation (\ref{fried})
$H^2/H_0^2$ is written as a function of the scale factor $a=\frac{1}{1+z}$ (or redshift $z$)
\begin{equation}
\frac{H^2}{H_0^2}=\frac{b^2}{\left(b+\frac{d}{2}\right)^2}\left(\Omega_{\gamma}\Omega_{\text{ch}}^2\frac{(K-3)(K+1)}{2b}+\Omega_{\text{ch}}
+\Omega_k\right),\label{hubble}
\end{equation}
where $\Omega_k=-\frac{k}{H_0^2 a^2}$.
%\begin{equation}
%\Omega_{\text{bm}}=\Omega_{\text{bm},0}a^{-3}=\frac{\rho_{\text{bm},0}}{3H_0^2}a^{-3}\;\;\;\text{and}\;\;\;
%\Omega_k=-\frac{k}{H_0^2 a^2}.
%\end{equation}
The function $b$ is positive as both $\Omega_\gamma$ and $\Omega_R$ are positive.

We can also add to formula (\ref{hubble}) dimensionless parameter $\Omega_{\text{r}}=\Omega_{\text{r,0}} a^{-4}=\frac{\rho_{\text{r}}}{3H_0^2}a^{-4}$ related with the radiation. Pressure of the radiation is equal $p_{\text{r}}=\frac{1}{3}\rho_{\text{r}}$, where $\rho_{\text{r}}$ is density of the radiation. Because trace of radiation is equal zero then the formula (\ref{palatinir}) for $\hat{R}$ is not changed. In consequence, formulas for $b$ and $d$ are the same. Expression (\ref{hubble}) for the case with radiation has the following form
 \begin{equation}
 \frac{H^2}{H_0^2}=\frac{b^2}{\left(b+\frac{d}{2}\right)^2}\left(\Omega_{\gamma}\Omega_{\text{ch}}^2\frac{(K-3)(K+1)}{2b}+\Omega_{\text{ch}}+\frac{\Omega_{\text{r}}}{b}
 +\Omega_k\right) .\label{hubble2}
 \end{equation}

Let us notice that if the coordinate transformation $t\rightarrow \tau \colon \frac{|b| dt}{|b+\frac{d}{2}|}=d\tau$ is non-singular, then the relation (\ref{hubble}) can be rewritten into the form
\begin{equation}
\frac{H^2(\tau)}{H_0^2}=\Omega_{\gamma}\Omega_{\text{ch}}^2\frac{(K-3)(K+1)}{2b}+\Omega_{\text{ch}}%+\Omega_{\text{bm}}
+\Omega_k ,
\end{equation}
where new Hubble parameter is $H(\tau)=a(\tau)^{-1}\frac{da(\tau)}{d\tau}$. Note that new time $\tau$ is growing function of the original cosmological time $t$.
In order to investigate whether the reparameterization of time is a diffeomorphism, let us consider zero of the function
\begin{equation} \label{eq:13}
f(K,\alpha,A_{\text{s}}, \Omega_{\gamma})=2b+d\Longrightarrow
\alpha K^2-3(1+\alpha)K-\frac{K^{\frac{1}{1+\alpha}}}{\Omega_{\gamma}\Omega_{\text{ch,0}}\left(3A_{\text{s}}\right)^{\frac{1}{1+\alpha}}}+1=0.
\end{equation}
The real solution of the above algebraic equation in the interval [0, 3) will determine the position of the
singularity $a_{\text{sing}}\text{: } f(K(a_{\text{sing}}))=0$. At this singular point the value $a=a_{\text{sing}}$ is
finite and can be reached at finite time $t_{\text{sing}}$. Instead one has $\dot{a}_{\text{sing}}=\infty$ and $|\ddot{a}_{\text{sing}}|=\infty$ (the vertical inflection point)
\footnote{It is rather known (see e.g. \cite{Borowiec:2011wd,Borowiec:2013ima}) that Palatini cosmological models are dynamical systems of Newtonian type: $\ddot{a}=-\frac{d V(a)}{da}$, where the (non-positive) potential is determined by the Hubble rate $V(a)=-\frac{1}{2}a^2 H^2$. Thus our singular point is just a pole of such potential function.}. In more physical terms one reads that an effective matter density $\rho_{\text{eff}}$ and an effective pressure $|p_{\text{eff}}|$ are singular (infinite). Such a singularity is called the freeze singularity or a singularity of type III
(see e.g \cite{Nojiri:2015wsa,Nojiri:2005sx} and references therein for classification and properties of various cosmological singularities).
Freeze singularities were previously studied in different context, also in relations with GCG, Loop Quantum Cosmology, etc..
(see e.g. \cite{Yurov,Lopez,Singh} and references therein). In our case it provides
decreasing comoving Hubble radius $(aH)^{-1}\rightarrow 0$ and hence can serve for inflationary epoch.
More detailed analysis will be presented in a subsequent publication. We remark that these type of singularities are  generic property of Palatini cosmological models \cite{Borowiec:2011wd,Borowiec:2013ima}.

Let us start with the special case $\alpha=1$, it means the original Chaplygin gas. Then the equation (\ref{eq:13}) assumes the following form
\begin{equation}
K^4-12K^3+38K^2-\chi K+1=0,
\end{equation}
where
\begin{equation}
\chi =\left(12+\frac{1}{3 A_{\text{s}}\Omega _{\gamma }^2\Omega _{\text{ch},0}^2}\right) \text{ and } \chi\in [12,\text{ }\infty).
\end{equation}
From our numerical analysis and the bound $K_{\text{sing}}\leq \frac{3}{1+(\frac{1-A_{\text{s}}}{A_{\text{{s}}}})a_{\text{sing}}^{-3(1+\alpha)}}$ we obtain that there is a single singularity of type III. It is interesting that zero of $f(K)$ function can be obtained in the exact form
\begin{equation}
K_{\text{sing}} = 3- \frac{\zeta}{\sqrt{6}} -
 \sqrt{\left( \frac{16}{3}+\frac{(9 \chi-364 )}{3\xi }-\frac{1}{12} \xi
 -\frac{\sqrt{6} (\chi-12 )}{4\zeta}\right)},
\end{equation}
where
\begin{align}
\zeta(\chi)&=\sqrt{16+\frac{2(364-9 \chi )}{\xi }  +\frac{\xi}{4} },\\
\xi(\chi) &= \left(55448-2052 \chi +\frac{27 \chi ^2}{2}+\frac{3}{2} \sqrt{3(\chi-12)^2 \left(27 \chi ^2-12496-648 \chi \right)}\right)^{1/3}.
\end{align}
Therefore, the position of the singularity crucially depends on the $\chi$ parameter which for $\Omega_{\gamma}\ll 1$ is $\chi=\left(3A_{\text{s}}\Omega_{\gamma }^2\Omega_{\text{ch},0}^2\right)^{-1}$
.
The scale factor for this case is expressed by the following formula
\begin{equation}
a_{\text{sing}}=\left(\frac{A_{\text{s}}}{1-A_{\text{s}}}\right)^{\frac{-1}{3(1+\alpha)}}
\left(\frac{3}{3- \frac{\zeta}{\sqrt{6}}- \sqrt{\left(\frac{16}{3}+\frac{(9 \chi-364 )}{3
			\xi }-\frac{1}{12} \xi -\frac{\sqrt{6} (\chi-12 )}{4\zeta}\right)}}-1\right)^{\frac{-1}{3(1+\alpha)}}  .
\end{equation}

If $\alpha=0$ then the matter content of our universe is the same as in $\Lambda$CDM model
\footnote{In fact, the case $\alpha=\gamma=0$ reconstructs $\Lambda$CDM model completely. However we should notice that the presence of the quadratic term ($\gamma\neq 0$) is crucial for our considerations. For $\gamma=0$ one gets $b=1, d=0$ and the equation (\ref{eq:13}) has no solutions at all. From the other hand the value $\gamma<<1$ should be very small in order to locate the singularity in an appropriate epoch.}. For this case the equation $f(K)=0$ possesses a solution if there
exists a value of the scale factor for the singularity. Expressions for $K_{\text{sing}}$ and the $a_{\text{sing}}$ are the following
\begin{equation}
K_{\text{sing}}=\frac{1}{3+\frac{1}{3\Omega_{\gamma}\Omega_{\text{ch,0}}A_{\text{s}}}}\;\;
\;\;\text{and}\;\;\;\;
a_{\text{sing}}=\left(\frac{1-A_{\text{s}}}{8A_{\text{s}}+\frac{1}{\Omega_{\gamma}\Omega_{\text{ch,0}}}}\right)^\frac{1}{{3}}\;\;\;
\text{for}\;\; \alpha=0 .
\end{equation}

Therefore, we should expect that in the case of homogeneous cosmological models with Chaplygin gas (which is reduced to the $\Lambda$ and dust matter) there is the freeze type of singularity.

In the general case we have that
\begin{equation}
\frac{b^2}{\left(b+\frac{d}{2}\right)^2}=\frac{\left(1+2\Omega_{\gamma}\Omega_{\text{ch,0}}(K+1)\right)^2}{\left(1+\Omega_{\gamma}\Omega_{\text{ch,0}}(3K+\alpha K(3-K)-1)\right)^2}.
\end{equation}

Note that all density parameters $\alpha$, $\Omega_{\text{ch},0}$,  %$\Omega_{\text{bm},0}$,
$\Omega_{\text{k},0}$, $A_{\text{s}}$, $\Omega_{\gamma}$ are not independent and satisfy the constraint condition
\begin{equation}
\begin{array}{l}
1 - \Omega_{\text{ch},0} - \Omega_{k,0} =\frac{\Omega_{\gamma}\Omega_{\text{ch,0}}}{2+4\Omega_{\gamma}\Omega_{\text{ch,0}}(3A_{\text{s}}+1)}\times\\ \\
\times(1-A_{\text{s}})(1-3\alpha A_{\text{s}})\left(12-3\Omega_{\text{ch,0}}+\frac{6\Omega_{\gamma}\Omega_{\text{ch,0}}}{1+2\Omega_{\gamma}\Omega_{\text{ch,0}}(3A_{\text{s}}+1)}\right)  .
\end{array}
\end{equation}

\section{Statistical analysis}

In this section a statistical analysis of the considered model is presented: we have used the SNIa, BAO, CMB and lensing observations, measurements of $H(z)$ for galaxies and the Alcock-Paczy{\'n}ski test.

The data from Union 2.1 which is the sample of 580 supernovae \cite{Suzuki:2011hu} has been also used. The likelihood function for SNIa is expressed by the
following formula
\begin{equation}
	\ln L_{\text{SNIa}} = -\frac{1}{2} [A - B^2/C + \log(C/(2 \pi))],
\end{equation}
where $A= (\mathbf{\mu}^{\text{obs}}-\mathbf{\mu}^{\text{th}})\mathbb{C}^{-1}(\mathbf{\mu}^{\text{obs}}-\mathbf{\mu}^{\text{th}})$,
$B= \mathbb{C}^{-1}(\mathbf{\mu}^{\text{obs}}-\mathbf{\mu}^{\text{th}})$, $C=\mathrm{tr} \mathbb{C}^{-1}$ and $\mathbb{C}$ is a covariance matrix for SNIa.
The distance modulus is expressed by $\mu^{\text{obs}}=m-M$ (where $m$ is the apparent magnitude and  $M$ is the absolute magnitude of SNIa)
and $\mu^{\text{th}} = 5 \log_{10} D_L +25$ (where the luminosity distance is given by $D_L= c(1+z) \int_{0}^{z} \frac{d z'}{H(z)}$).

We have used Sloan Digital Sky Survey Release 7 (SDSS DR7) dataset for $z = 0.275$ \cite{Percival:2009xn}, 6dF Galaxy Redshift Survey measurements for redshift $z = 0.1$ \cite{Beutler:2011hx}, the BOSS DR 9 measurements for $z = 0.57$ \cite{Anderson:2012sa,Blake:2011ep}, and WiggleZ measurements for redshift $z = 0.44, 0.60, 0.73$ \cite{Blake:2012pj}.
The likelihood function for BAO has the following form
\begin{equation}
	\ln L_{\text{BAO}} = - \frac{1}{2}\left(\mathbf{d}^{\text{obs}}-\frac{r_s(z_d)}{D_V(\mathbf{z})}\right)\mathbb{C}^{-1}\left(\mathbf{d}^{\text{obs}}
	-\frac{r_s(z_d)}{D_V(\mathbf{z})}\right),
\end{equation}
where $r_s(z_d)$ is the sound horizon at the drag epoch \cite{Eisenstein:1997ik}.

The likelihood function for the Planck observations of cosmic microwave background radiation (CMB) \cite{Ade:2013zuv}, lensing, and low-$\ell$ polarization from the WMAP (WP) is given by the following expression
\begin{equation}
	\ln L_{\text{CMB}+\text{lensing}+\text{WP}} = - \frac{1}{2}  (\mathbf{x}^{\text{th}}-\mathbf{x}^{\text{obs}})
	\mathbb{C}^{-1} (\mathbf{x}^{\text{th}}-\mathbf{x}^{\text{obs}}),
\end{equation}
where $\mathbb{C}$ is the covariance matrix with the errors, $\mathbf{x}$ is a vector of the acoustic scale $l_{A}$, the shift parameter
$R$ and $\Omega_{b}h^2$ where
\begin{align}
	l_A &= \frac{\pi}{r_s(z^{*})} c \int_{0}^{z^{*}} \frac{dz'}{H(z')} ,\\
	R &= \sqrt{\Omega_{\text{m,0}} H_0^2} \int_{0}^{z^{*}} \frac{dz'}{H(z')} ,
\end{align}
where $z^{*}$ is the redshift of the epoch of the recombination.

The likelihood function for the Alcock-Paczynski test \cite{Alcock:1979mp,Lopez-Corredoira:2013lca} is given by the following formula
\begin{equation}
	\ln L_{AP} =  - \frac{1}{2} \sum_i \frac{\left( AP^{th}(z_i)-AP^{obs}(z_i) \right)^2}{\sigma^2}.
\end{equation}
where $AP(z)^{\text{th}} \equiv \frac{H(z)}{z} \int_{0}^{z} \frac{dz'}{H(z')}$ and $AP(z_i)^{\text{obs}}$ are observational
data \cite{Sutter:2012tf,Marinoni:2010yoa,Ross:2006me,daAngela:2005gk,Outram:2003ew,Paris:2012iw,Schneider:2010hm}.

We are using data of $H(z)$ for galaxies from \cite{Simon:2004tf,Stern:2009ep,Moresco:2012jh}. The likelihood function is expressed by
\begin{equation}\label{hz}
	\ln L_{H(z)} = -\frac{1}{2} \sum_{i=1}^{N}  \left (\frac{H(z_i)^{\text{obs}}-H(z_i)^{\text{th}}}{\sigma_i }\right)^2.
\end{equation}

The final likelihood function is
\begin{equation}
	L_{\text{tot}} = L_{\text{SNIa}} L_{\text{BAO}} L_{\text{AP}} L_{H(z)}.
\end{equation}

We used our own code CosmoDarkBox in estimation of the model parameters. This code uses the Metropolis-Hastings algorithm \cite{Metropolis:1953am,Hastings:1970aa}.

\begin{figure}%[ht]
	\centering
	\includegraphics[scale=1]{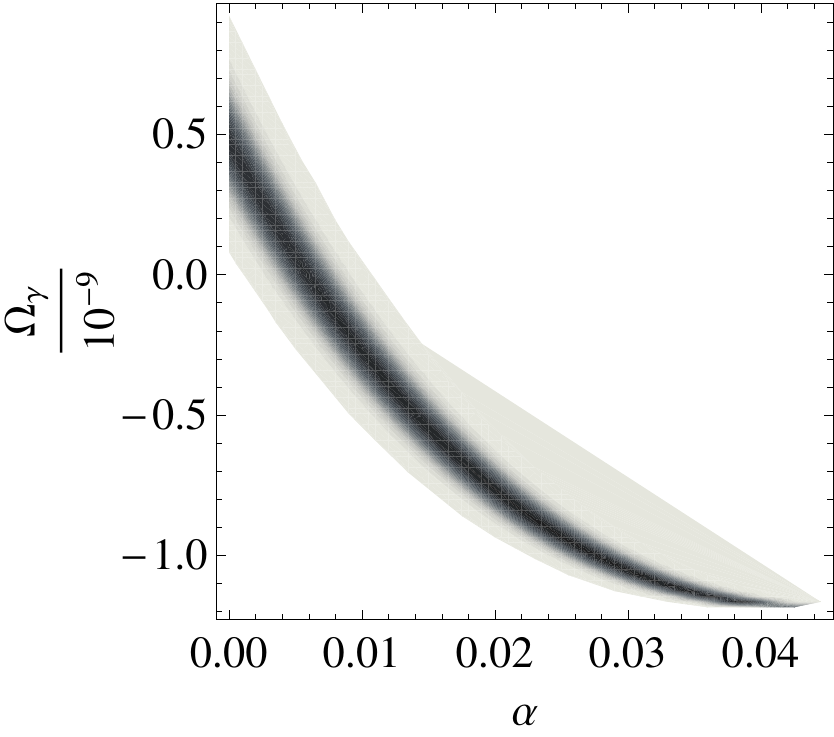}
	\caption{The likelihood function of two model parameters ($\alpha$, $\Omega_{\gamma}$) with the marked $68\%$ and $95\%$ confidence levels for model with the Chaplygin gas and radiation. We assume $H_0=67.27\frac{\text{km}}{\text{s Mpc}}$, $A_\text{s}=0.6908$.}
	\label{fig:6}
\end{figure}

\begin{figure}%[ht]
	\centering
	\includegraphics[scale=1]{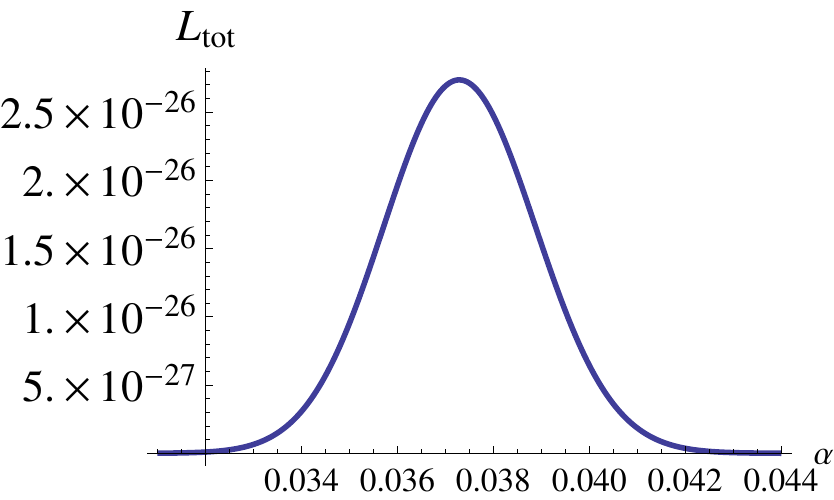}
	\caption{Diagram of PDF for parameter
		$\alpha$ obtained as an intersection of a likelihood function for model with the Chaplygin gas and radiation. Two planes of intersection likelihood function are
		$\Omega_{\gamma}=-1.15\times 10^{-9}$ and $A_{\text{s}}=0.6908$.}
	\label{fig:7}
\end{figure}

\begin{figure}%[ht]
	\centering
	\includegraphics[scale=1]{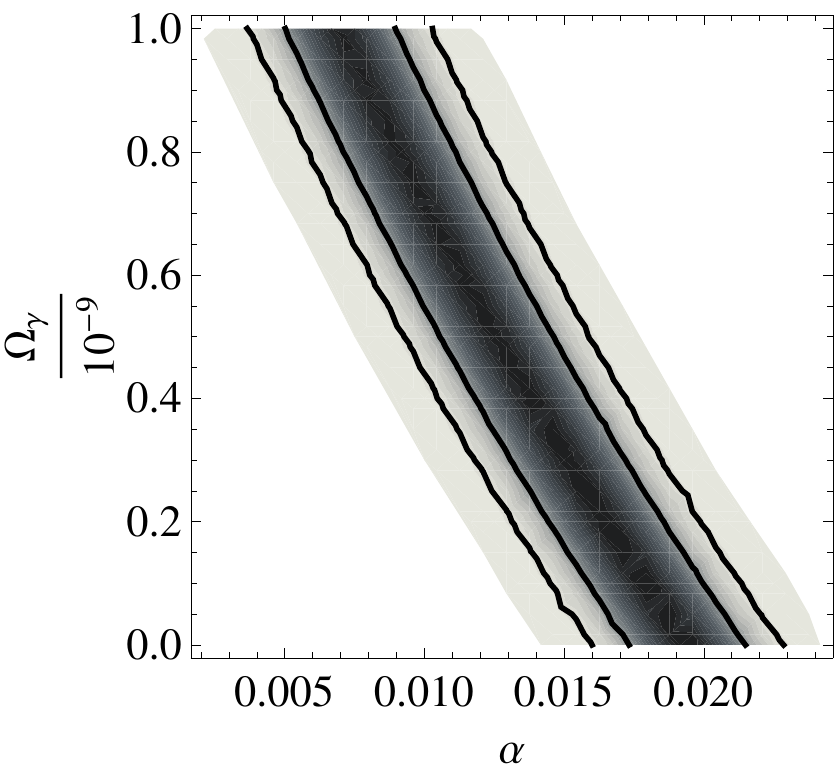}
	\caption{The likelihood function of two model parameters ($\alpha$, $\Omega_{\gamma}$) with the marked $68\%$ and $95\%$ confidence levels for model with the Chaplygin gas and barionic matter. We assume $H_0=67.27\frac{\text{km}}{\text{s Mpc}}$, $A_\text{s}=0.7264$.}
	\label{fig:1}
\end{figure}

\begin{figure}%[ht]
	\centering
	\includegraphics[scale=1]{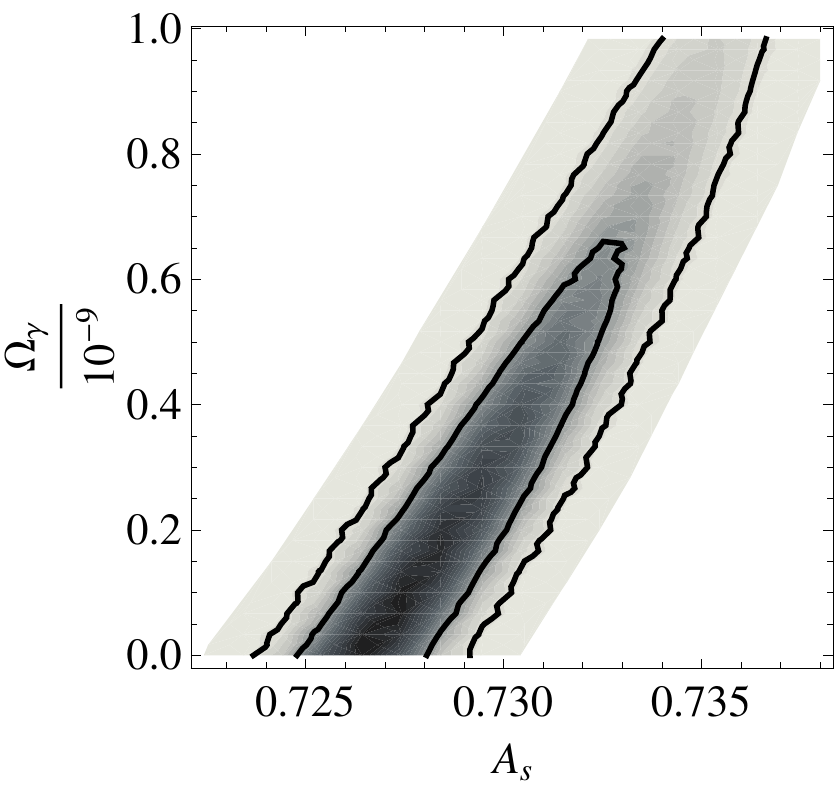}
	\caption{The likelihood function of two model parameters ($A_{\text{s}}$, $\Omega_{\gamma}$) with the marked $68\%$ and $95\%$ confidence levels for model with the Chaplygin gas and barionic matter. We assume $H_0=67.27\frac{\text{km}}{\text{s Mpc}}$, $\alpha=0.0194$.}
	\label{fig:2}
\end{figure}

\begin{figure}%[ht]
	\centering
	\includegraphics[scale=1]{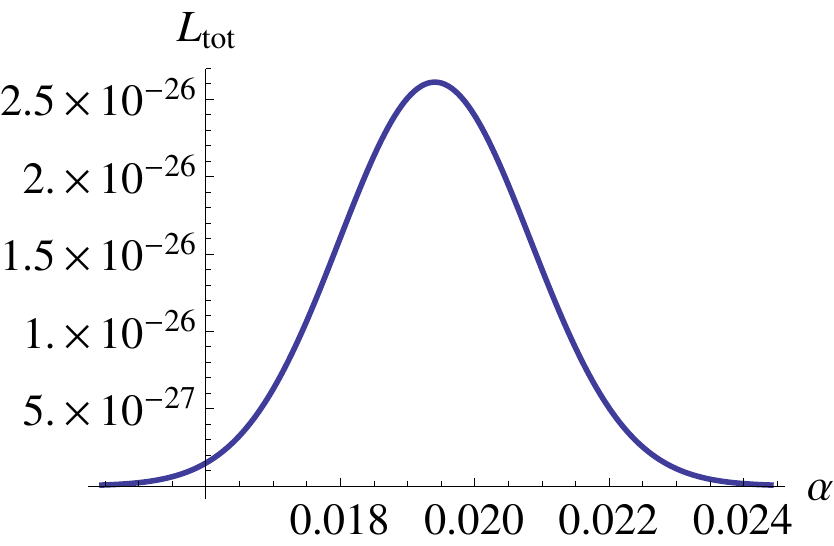}
	\caption{Diagram of PDF for parameter
	$\alpha$ obtained as an intersection of a likelihood function for model with the Chaplygin gas and barionic matter. Two planes of intersection likelihood function are
	$\Omega_{\gamma}=0$ and $A_{\text{s}}=0.7264$.}
	\label{fig:3}
\end{figure}

\begin{figure}%[ht]
	\centering
	\includegraphics[scale=1]{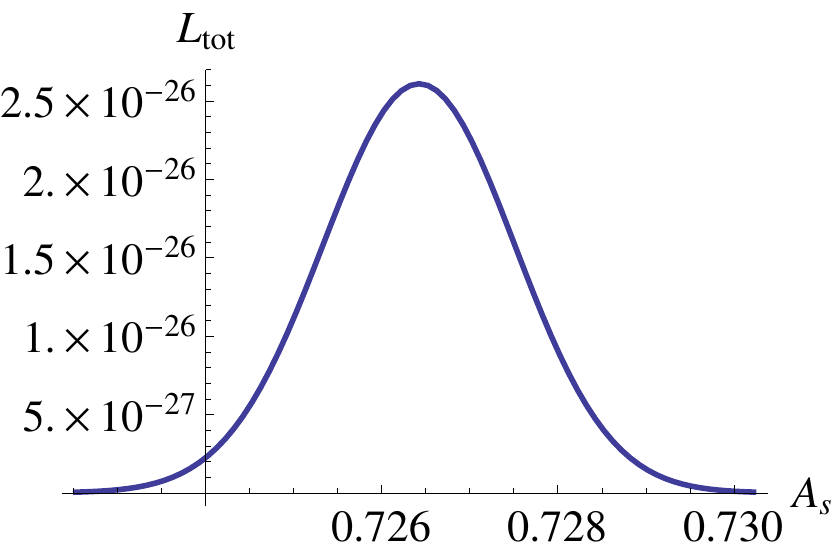}
	\caption{Diagram of PDF for parameter
		$A_{\text{s}}$ obtained as an intersection of a likelihood function for model with the Chaplygin gas and barionic matter. Two planes of intersection likelihood function are
		$\Omega_{\gamma}=0$ and $\alpha=0.0194$.}
	\label{fig:4}
\end{figure}

In model estimation we assume that model possesses three free parameters: $A_s, \alpha$ and $\Omega_\gamma$. It is assumed for simplicity that model is flat and parameter $\Omega_{\text{ch}}$ is derived from the corresponding constraint condition. Accordingly, we also assume that  $H_0=67.27\frac{\text{km}}{\text{s Mpc}}$ and redshift of matter-radiation
equality $z_{\text{eq}}=3395$ and the value of $\Omega_{\gamma}<10^{-9}$. If the value of $\Omega_{\gamma}$ had been more than $10^{-9}$ then the epoch of the freeze singularity would have been in the epoch of recombination or after. We consider two models. The first model is for the Chaplygin gas with radiation (see equation (\ref{hubble2})). The second model is for the Chaplygin gas with barionic matter which is described by formula
\begin{equation}
\frac{H^2}{H_0^2}=\frac{b^2}{\left(b+\frac{d}{2}\right)^2}\left(\Omega_{\gamma}\Omega_{\text{ch}}^2\frac{(K-3)(K+1)}{2b}+\Omega_{\text{ch}}+\Omega_{\text{bm}}
+\Omega_k\right),
\end{equation}
where $\Omega_{\text{bm}}=\Omega_{\text{bm},0}a^{-3}$ related with the presence of baryonic visible matter for which the parameter $\Omega_{\text{bm},0}=0.04917$ is assumed following the Planck estimation. The results of statistical analysis for the first model are represented in Tables~\ref{table:1} and \ref{table:3} and in figure~\ref{fig:6}, where it is shown the likelihood function with $68\%$ and $95\%$ confidence level. The results for the second model are shown in Table~\ref{table:2} and in figures~\ref{fig:1}, \ref{fig:2}. PDF diagrams are presented in figures \ref{fig:7}, \ref{fig:3} and \ref{fig:4}.

\begin{table}
	\caption{The best fit and errors for model with the Chaplygin gas and radiation for the case where we assume the value of $\Omega_{\gamma}$ from the
		interval $(-1.2\times10^{-9}, 10^{-9})$. We assume also $A_{\text{s}}$ from the interval $(0.67, 0.72)$, and $\alpha$ from the interval $(0, 0.06)$. The value of $\chi^2$ for the best fit is equaled 117.722.}
	\label{table:1}
	\begin{center}
		\begin{tabular}{llll} \hline
			parameter & best fit & $ 68\% $ CL & $ 95\% $ CL\\
			\hline \hline
			$A_{\text{s}}$ & 0.6908 & $\begin{array}{c} +0.0066
			\\ -0.0069
			\end{array}$ & $\begin{array}{c} +0.0104
			 \\ -0.0098
			\end{array}$
			\\ \hline
			$\alpha$ & 0.0373 & $\begin{array}{c} +0.0083
			\\ -0.0373
			\end{array}$ & $\begin{array}{c} +0.0131
			\\ -0.0373
			\end{array}$ \\ \hline
			$\Omega_{\gamma}$ & $-1.156\times 10^{-9}$ & $\begin{array}{c} +2.156\times 10^{-9}
			\\ -0.010\times 10^{-9}
			\end{array}$ & $\begin{array}{c} +2.156\times 10^{-9}
			\\ -0.015\times 10^{-9}
			\end{array}$
			\\ \hline
		\end{tabular}
	\end{center}
\end{table}

\begin{table}
	\caption{The best fit and errors for model with the Chaplygin gas and radiation for the case where we assume the value of $\Omega_{\gamma}$ from the
		interval $(-1.2\times10^{-9}, 0)$. We assume also $A_{\text{s}}$ from the interval $(0.67, 0.72)$, and $\alpha$ from the interval $(0, 0.06)$. The value of $\chi^2$ for the best fit is equaled 117.722.}
	\label{table:3}
	\begin{center}
		\begin{tabular}{lllll} \hline
			parameter & best fit & $ 68\% $ CL & $ 95\% $ CL\\
			\hline \hline
			$A_{\text{s}}$ & 0.6908
			& $\begin{array}{c} +0.0065
			\\ -0.0068
			\end{array}$ & $\begin{array}{c} +0.0103
			\\ -0.0098
			\end{array}$
			\\ \hline
			$\alpha$ & 0.0373
			& $\begin{array}{c} +0.0080
			\\ -0.0373
			\end{array}$ & $\begin{array}{c} +0.0129
			\\ -0.0373
			\end{array}$\\ \hline
			$\Omega_{\gamma}$ & $-1.156\times 10^{-9}$
			& $\begin{array}{c} +1.156\times 10^{-9}
			\\ -0.008\times 10^{-9}
			\end{array}$ & $\begin{array}{c} +1.156\times 10^{-9}
			\\ -0.014\times 10^{-9}
			\end{array}$
			\\ \hline
		\end{tabular}
	\end{center}
\end{table}

\begin{table}
	\caption{The best fit and errors for model with Chaplygin gas and barionic matter. We assume $A_{\text{s}}$ from the interval $(0.70, 0.74)$, $\Omega_{\gamma}$ from the
	interval $(0, 10^{-9})$ and $\alpha$ from the interval $(0, 0.04)$. $\Omega_{\text{bm},0}$ is assumed to be 0.04917. The best fit of $\Omega_{\gamma}$ is
	zero. The value of $\chi^2$ for the best fit is equaled 117.816.}
	\label{table:2}
	\begin{center}
		\begin{tabular}{llll} \hline
			parameter & best fit & $ 68\% $ CL & $ 95\% $ CL  \\ \hline \hline
			$A_{\text{s}}$ & 0.7264 & $\begin{array}{c}
			+ 0.0068\\ -0.0069
			\end{array}$ & $\begin{array}{c}
			+0.0110 \\ -0.0113
			\end{array}$\\ \hline
			$\alpha$ & 0.0194 & $\begin{array}{c}
			+ 0.0090\\ -0.0194
			\end{array}$ & $\begin{array}{c}
			+ 0.0146\\ -0.0194
			\end{array}$ \\ \hline
		\end{tabular}
	\end{center}
\end{table}

The value of $\chi^2$ for the best fit for the model with the Chaplygin gas and barionic matter is equaled to 117.816 and the value of reduced $\chi^2$ is equal 0.1894. For the model with the Chaplygin gas and radiation $\chi^2$ for the best fit is equaled 117.722 and the value of reduced $\chi^2$ is equal 0.1892. We have used the BIC (Bayesian Information Criterion). The BIC is defined as follows \cite{Schwarz:1978ed,Kass:1995bf}
\begin{equation}
\text{BIC}= \chi^2 \,+\, j\, \ln(n),
\end{equation}
where $j$ is number of parameters and $n$ is the number of data points. In this paper, $n=625$. Although number of parameters for our model is
6 ($H_0$, $\Omega_{\text{bm},0}$, $\Omega_{\text{r},0}$, $\Omega_{\gamma}$, $A_{\text{s}}$, $\alpha$) we took $j=3$ in computation of BIC. It is so since we assumed
some values for $H_0$, $\Omega_{\text{b},0}$ and $\Omega_{\text{r},0}$ in the estimation. The value of BIC for our
model with radiation is 137.036. For comparison: BIC of $\Lambda$CDM model is 125.303 (the value of $\chi^2$ is 118.866 while reduced $\chi^2$ is 0.1908). For computation of BIC
of $\Lambda$CDM model we took $j=1$ because, as previously, we assumed that the values of $H_0$, $\Omega_{\text{b},0}$ and $\Omega_{\text{r},0}$ are already known.
In consequence, the only free parameter is $\Omega_{\text{m},0}$ representing matter. The difference between BIC of our model and $\Lambda$CDM model is $\Delta\text{BIC}=11.733$. If the value $\Delta$BIC is
between the numbers 2 and 6 then the evidence againt the model is positive in comparison to the model of the null hypothesis. If that value is more than 6, the evidence against the model is strong \cite{Kass:1995bf}. Consequently, the evidence in favor  $\Lambda$CDM model is strong in comparison to our model.

From our statistical analysis we obtain that the negative best value of $\Omega_{\gamma}$ is favored. This means that models with bounce instead of the initial singularity are favored. The values of $\chi^2$ for our model and $\Lambda$CDM model are comparable.
\section{Conclusions}\label{conclus}

The aim for the paper was to insert inflationary era  as an internal gravitational process into the evolution of early universe. We were motivated by recent papers \cite{Nojiri:2015wsa,Nojiri:2015fra,Barrow:2015,Nojiri:2015fia,Odintsov:2015tka,Odintsov:2015zza,Nojiri:2005sx}, mostly phenomenological ones, relating type IV singularities with the inflation. Our idea was to construct concrete cosmological model which could provide on the one hand similar inflationary scenario  and reach a good agreement with the present day experimental data from the other hand.
%\cite{Nojiri:2015wsa,Odintsov:2015zza,Nojiri:2015fra,Nojiri:2005sx,Nojiri:2015fia,Odintsov:2015tka} this
%type of singularities is related with the inflation.
For checking this hypothesis we considered cosmological model with Chaplygin gas in the Palatini
formalism where freeze singularities of type III do appear in a natural way as poles in a Newtonian potential \cite{Borowiec:2011wd,Borowiec:2013ima}.
Inserting this type of singularity into the evolution of early universe provides the inflation in a sense that the corresponding comoving Hubble radius decreases: $(aH)^{-1}\rightarrow 0$ approaching the singularity.
We choose the parametrization of gravity as $f(\hat{R}) = \hat{R} + \gamma \hat{R}^2$ in order to accomplish minimal deviation from Einstein gravity.
Additionally, it was speculated that the singularity happens before the recombination epoch which guarantees the preservation of post-recombination physics of the Universe. From this assumption we obtained an upper limit on the value of the parameter $\Omega_{\gamma}$. We also assumed that this parameter is positive at the very beginning because in the early universe when the Ricci scalar $R \gg 1$, the term $R^2$ will dominate over $R$ in the Lagrangian. Therefore, $\Omega_{\gamma}$ is defined in the interval $[0; 10^{-9})$.

Our statistical analysis shows that the cosmological model is in a good agreement
with observational data. The model under consideration provides additional density parameter $\Omega_\gamma$ which measures the deviation from the Chaplygin gas model. Let us notice that our model possesses three parameters to fit. We obtained a best fit and confidence intervals for three parameters $\alpha$, $A_s$ and $\Omega_{\gamma}$ which are consistent with other estimations beyond the context of the Palatini formalism. % \textcolor{green}{cytacja}.
From the other hand the only very small (but nonzero) values of $\gamma$ are allowed in order to locate singularity before recombination epoch. From statistical analysis we obtain that the negative value of $\Omega_{\gamma}$ is favored.

We compared the $\Lambda$CDM model with our model using BIC (Bayesian information criterion). From this comparison $\Delta$BIC=11.733. In consequence the evidence in favor $\Lambda$CDM model is strong in comparison to our model, while the value of reduced  $\chi^2$  is in favor of new models. We have found that the presented modified model of gravity admitted III type singularity in the early time of the universe evolution. Furthermore, the redshift corresponding to this singularity for the early universe was also computed. In addition, we shortly discussed the special cases of GCG, it means $\alpha$=0, $\alpha$=1 and for them we also recovered the exact expressions for the value of redshift corresponding to freeze singularity.

Concluding, the investigation of the model gave us the result in a form of the four phases of the cosmic evolution: the decelerating phase dominated by matter, an intermediate inflationary phase corresponding to III type singularity, a phase of matter domination (decelerating phase) and finally, the phase of acceleration of current universe. Big Bang singularity is preserved due
to the presence of radiation/barionic dust matter term incorporated into the Friedmann equation. In the case of negative $\gamma<0$ a bounce instead of Big
Bang is more favorable. It should be also remarked that a presence of freeze singularity is a generic feature of models in the Palatini formalism.
%\textcolor{green}{Co jeszcze na ten temat mowia prace dotyczace gazu chaplygina - czy tam tez pojawiaja sie tego typu osobliwosci.}

\acknowledgments{
The work was supported by the NCN project DEC-2013/09/B/ST2/03455.  AW acknowledges financial
support from 1445/M/IFT/15. The first author acknowledges hospitality from the
%Department of Sciences and Technological Innovation, University of Easter Piedmont and the
Department of Mathematics of Turin University, and personally of Lorenzo Fatibene,  during the preparation of the manuscript.}

%\bibliography{astrophysics,astrophysmath}
\bibliographystyle{JHEP}

\providecommand{\href}[2]{#2}\begingroup\raggedright\endgroup

\end{document}